\documentclass[12pt,a4paper]{article}  

\usepackage{snapshot}
\usepackage[T1]{fontenc}
\usepackage[utf8]{inputenc}
\usepackage{a4wide}

\usepackage{calc}
\usepackage{graphicx}
\usepackage{macromath}
\usepackage{mythm}
\usepackage{esint}
\usepackage[dvips]{hyperref}
\usepackage[all]{xy}

\newcommand{\fib}[1]{#1}
\renewcommand{\sec}[1]{\mathcal{#1}}
\newcommand{\alg}[1]{\mathcal{#1}}

\newcommand{\fE}{\fib{E}}
\newcommand{\sE}{\sec{E}}
\newcommand{\fP}{\fib{P}}
\newcommand{\sF}{\sec{F}}
\newcommand{\sM}{\sec{M}}

\newcommand{\aA}{\alg{A}}
\newcommand{\aB}{\alg{B}}

\begin{document}

\author{Emmanuel Sérié}
\title{Noncommutative gauge theories with Endomorphism algebras}
\date{\today}

\maketitle
\begin{center}
  Laboratoire de Physique Théorique (UMR 8627)\\
  Université Paris XI,\\
  Bâtiment 210, 91405 Orsay Cedex, France
\end{center}
\begin{abstract}
  
We formulate a Yang-Mills action principle for noncommutative connections on an endomorphism algebra of a vector bundle.
It is shown that there is an influence of the topology of the vector bundle onto 
the structure of the vacuums of the theory in a non common way.
This model displays a new kind of symmetry breaking mechanism.
Some mathematical tools are developed in relation with endomorphism algebras and a new approach of the usual Chern-Weil homomorphism in topology is given.
\end{abstract}
\vfill

LPT-Orsay 06-53
\newpage
\tableofcontents

\newpage
\section*{Introduction}
 
We generalize the theory of noncommutative connections and noncommutative Yang-Mills action developed in~\cite{dubois-violette:89,dubois-violette:89:II,dubois-violette:90:II} on the algebra $C^{\infty}(M)\otimes M_{n}(\gC)$ to the situation of non trivial fiber bundles, \textit{i.e.}  the situation where the algebra  $C^{\infty}(M)\otimes M_{n}(\gC)$ is replaced by the endomorphism algebra of a non trivial vector bundle.
The present work is a direct continuation of the work made in \cite{dubois-violette:98,masson:99} where the key notions of the noncommutative geometry of endomorphism algebras where introduced.
The present article contain also non-published results obtained in \cite{serie-phd:05}.

The goal of the three first sections is to define the mathematical notions necessary to write and define correctly a Yang-Mills action principle for a noncommutative connection on a projective module over an endomorphism algebra.

In the first section, we give an introduction to the notion of endomorphism algebra and we illustrate the idea initiated in \cite{masson:99} that all notions which are defined in the framework of principal fiber bundles can be formulated in a completely algebraic language using endomorphism algebras.
We introduce the notion of associated projective module corresponding to the notion of associated vector bundle and we generalize the notion of tensorial form.
At the end of this sectiion, we develop a new approach of the Chern-Weil homomorphism using endomorphism algebras.

In the second section, we introduce the notion of noncommutative connection which is essential to the formulation of a Yang-Mills action principle.
We see that ordinary connections play a particular role and are useful to decompose the degrees of freedom of a noncommutative connection in a ``covariant'' way.

In the third part, we introduce the notion of metric and of integration for endomorphism algebras.

In the last part, we formulate the Yang-Mills action principle and we perform an analysis of the vacuums of the theory.
We compare the results with the trivial situation and observe that the topology of the vector bundle can modify the structure of this vacuums in a very particular way.

In the appendix, we develop the notion of Levi-Cità connection in relation to the notion of metric introduced in section~\ref{sec:integr-riem-struct}.

\section{Endomorphism algebra \textit{v.s.} principal fiber bundle}

\subsection{Endomorphism algebras}\label{sec:endom-algebr}
\subsubsection{Definition}\label{sec:endom-algebr-definition}
We consider a vector bundle $\fE$ over a smooth manifold $M$ which is associated to a principal fiber bundle $P$ with structure group $SL(n)$ (or $SU(n)$).

We denote $\sE=\Gamma(\fE)$ the space of smooth section of $\fE$.
This space is isomorphic to the space of equivariant maps from $\fP$ to $V^{\fE}$, the vector space on which $\fE$ is modeled.
The group $SL(n)$ acts on $V^{\fE}$ with a representation $R^{\fE}$.
We will denote $\fE^{*}$ the dual vector bundle of $\fE$ and $\sE^{*}$ its space of sections. 
Then the bundle of endomorphisms of $\fE$ is  $\End(\fE)\simeq \fE \otimes \fE^{*}$.

The space of sections of $\End(\fE)$ is an algebra that we will denote $\aA^{\fE}$.
We will call $\aA^{\fE}$ the endomorphism algebra of $\fE$ and we have:
\begin{align*}
  \aA^{\fE}=\Gamma(\End(\fE))\simeq \Gamma(\fE\otimes \fE^{*}) \ .
\end{align*}

\subsubsection{Properties}

The center $\cZ(\aA^{E})$ of the algebra $\aA^{E}$ is isomorphic to $C^{\infty}(M)$. 
The isomorphism is given by the embedding:
\begin{align*}
  C^{\infty}(M) &\to \aA^{\fE} \\
  f & \mapsto f\cdot \gone 
\end{align*}
We can also define the maps $\tr$ and $\det$ from $\aA^{\fE}$ to $C^{\infty}(M)$ which generalize the trace and determinant maps on matrices.
On has then a natural splitting of the algebra  $\aA^{\fE}$ as a $\cZ(\aA^{\fE})$-module: 
\begin{align*}
  \aA^{\fE}=\aA^{\fE}_{0}\oplus \cZ(\aA^{\fE})
\end{align*}
where $\aA^{\fE}_{0}=\ksl(\aA^{\fE})$ is the sub-$\cZ(\aA^{\fE})$-module of elements of  $\aA^{\fE}$ without trace. 
It has naturally a structure of Lie algebra, with the commutator of the algebra as Lie bracket.

The algebra $\aA^{\fE}$ is Morita equivalent to the algebra $C^{\infty}(M)$. 
This equivalence can be explicitely shown by considering the $C^{\infty}(M)-\aA^{\fE}$-bimodule $\sE^{*}$ and the $\aA^{\fE}-C^{\infty}(M)$-bimodule $\sE$.
Then one has:
\begin{align*}
  & \aA^{\fE} \simeq \sE \otimes_{C^{\infty}(M)}\sE^{*} &
&  \text{and}&
  & C^{\infty}(M)\simeq \sE^{*}\otimes_{\aA^{\fE}} \sE
\end{align*}
Hence, we have an equivalence between the categories of (right) modules $\mathbf{M}_{\aA^{\fE}}$ and $\mathbf{M}_{C^{\infty}(M)}$.
As a consequence, we have that every right (resp. left) $\aA^{\fE}$-module $\sM$ is isomorphic to a $\sF\otimes_{C^{\infty}(M)}\sE^{*}$ (resp. $\sE\otimes_{C^{\infty}(M)}\sF$) with $\sF$ a module over $C^{\infty}(M)$.

\subsection{The algebra corresponding to a principal fiber bundle}
For a given principal fiber bundle $P$ with structure group $SL(n)$, we can consider the associated vector bundle $\fE=P \ltimes_{SL(n)} \gC^{n}$ corresponding to the fundamental representation of $SL(n)$ on $\gC^{n}$.
We will consider the algebra of endomorphism which correspond to this vector bundle as the algebra of endomorphism canonically associated to the principal fiber bundle $P$. We will denote it $\aA$.

We will see that this algebra  is a better candidate than the algebra $C^{\infty}(P)$ to represent in the category of algebras the principal fiber bundle $P$. Obviously, the fact that $P$ has a non abelian structure group is encoded in the noncommutativity of the algebra $\aA$.
In fact, it can be shown~\cite{masson:99} that $\aA$ and $P$ are related by the following relation:
\begin{align*}
  &  \aA\simeq  \left(C^{\infty}(P)\otimes M_{n}(\gC)\right)_{SL(n)-invariant} 
\simeq \Gamma(P\ltimes_{SL(n)}M_{n}(\gC))
\ .
\end{align*}
with the $SL(n)$ action given by the right multiplication on $P$ and the adjoint action on $M_{n}(\gC)$.
It is useful to introduce the bigger algebra $\aB=C^{\infty}(P)\otimes M_{n}(\gC)$. So $\aA$ is an invariant subalgebra of $\aB$.
We can remark that the algebra $C^{\infty}(P)$ is also an invariant subalgebra of $\aB$ but with a different action (just on matrices) of $SL(n)$.

\subsection{Derivations of the algebra}
We will now analyze the structure of derivations of the algebra $\aA$.
We have the canonic short exact sequence of Lie algebras and $\cZ(\aA)$-modules:
\begin{align}
  \xymatrix@R=0pt@C=20pt{ 
    \Int(\aA) \ar@{^(->}[r] & \der(\aA) \ar@{->>}^-{\rho}[r] & \Out(\aA)
  }\label{int-der-exact-sequence}
\end{align}
where $\der(\aA)$ is the space of derivations , $ \Int(\aA)$ is the space of inner derivations and $\Out(\aA)\simeq \der(\aA)/\Int(\aA)$ is the space of outer derivations of the algebra.
Here, one has $\Out(\aA)=\Gamma(TM)$, the Lie algebra of vector fields on $M$ and the map $\rho$ is induced by the restriction of derivations to the center of the algebra.

Inner derivations are in the image of the adjoint map:
\begin{align*}
\xymatrix@R=0pt@C=20pt{
  ad: & \aA \ar[r] & \Int(\aA)\\
  & \gamma \ar@{|->}[r] & ad_{\gamma} \makebox[0pt][l]{$: a \mapsto [\gamma,a]$}
}
\end{align*}
The restriction of this map to $\aA_{0}=\ksl(\aA)$ is an isomorphism of Lie algebras and $C^{\infty}(M)$-modules with inverse  given by:
\begin{align*}
  \xymatrix@R=0pt@C=20pt{
    i\theta:  
    & \Int(\aA) \ar^-{\simeq}[r] 
    & \aA_{0} \\ 
    & ad_\gamma \ar@{|->}[r] 
    & \gamma}
\end{align*}


The exact sequence~(\ref{int-der-exact-sequence}) reads then here:
\begin{equation}
\begin{aligned}
  \xymatrix@R=0pt@C=20pt{ 
    0 \ar[r]
    & \aA_{0} \ar^-{ad}[r] 
    & {\der(\aA)} \ar^-{\rho}[r] 
    & {\Gamma(TM)} \ar[r]
    & 0 \\
    &
    &  \cX        \ar@{|->}[r]   
    & \rho(\cX)       \\
    &
    \gamma  \ar@{|->}[r] 
    & \ad_{\gamma} 
    &
    }
\end{aligned}
  \label{der-exact-sequence}
\end{equation}

\subsection{Derivation based differential calculus}
There is a natural differential calculus on $\aA$ which is the differential calculus based on derivations introduced in \cite{dubois:88}.
This calculus is a direct generalization of the de~Rham differential calculus on the smooth manifold $M$, where vector fields are replaced by derivations of the algebra.

We first define the differential calculus $(\uOder(\aA),\hd)$, where
\begin{align*}
  \uOder^{*}(\aA)= \biggl(\bigwedge^{*}_{\cZ(\aA)}\der(\aA)\biggl)^{\star \aA}
\end{align*}
is the set of $\cZ(\aA)$-multilinear anti-symmetric applications from $\der(\aA)$ to $\aA$, and $\hd$ the differential define by the Koszul formula:
\begin{multline}
  \hd\omega(\cX_1, \dots , \cX_{n+1}) = \sum_{i=1}^{n+1} (-1)^{i+1} \cX_i  \omega( \cX_1, \dots \omi{i} \dots, \cX_{n+1}) \\
  + \sum_{1\leq i < j \leq n+1} (-1)^{i+j} \omega( [\cX_i, \cX_j], \dots \omi{i} \dots \omi{j} \dots , \cX_{n+1}) \ .
  \label{differential}
\end{multline}
for all $\omega \in \uOder^{n}(\aA)$.

It is also natural to define the differential calculus $(\Oder(\aA),\hd)$ where $\Oder(\aA)$ is the sub-graded differential algebra of $\uOder(\aA)$ generate by $\aA$.


For an endomorphism algebra, it can be show that $\uOder(\aA)\simeq\Oder(\aA)$.

\subsection{The role of ordinary connections}
\label{sec:role-ordin-conn}
\subsubsection{Splitting of exact sequence}
\label{sec:splitt-exact-sequ}
One can split the exact sequence of derivations~(\ref{der-exact-sequence}) as an exact sequence of $\cZ(\aA)$-modules with the help of an ordinary connection on $\fE$.
Let us consider a connection  $\nabla^\fE$ on $\fE$. 
The corresponding connection $\nabla=\nabla^{\fE}\otimes\nabla^{\fE^{*}}$ on $\End(\fE)$ is identified canonically whith an application of $\cZ(\aA)$-modules from $\Gamma(TM)$ into $\der(\aA)$.  
Then, the map $\nabla\circ \rho$ is a projector in $\End(\der(\aA))$ and it split the space $\der(\aA)$ into an horizontal part $(\der(\aA))_{\hor}\simeq \Gamma(TM)$ and a vertical part $(\der(\aA))_{\ver}\simeq \Int(\aA)\simeq \aA_{0}$.
Finally, one defines a $\cZ(\aA)$-module homomorphism $\alpha$ from $\der(\aA)$ to $\aA_{0}$ by setting:
\begin{align*}
  \alpha&= -i\theta \circ \left(\Id_{\der(\aA)}- \nabla\circ \rho\right) \ ,
\end{align*}
which defines an element of $\Oder^{1}(\aA)$. 
This expression makes sense because the projector $\Id_{\der(\aA)}- \nabla\circ \rho$ has its image in $\ker(\rho) \simeq \Int(\aA)$.
This construction can be sumarized by the following diagram:
\begin{align}
  \xymatrix@R=0pt@C=20pt{ 
    \aA_{0} 
    & \ar_-{-\alpha}[l]  {\der(\aA)} 
    & \ar_-{\nabla}[l] {\Gamma(TM)}\\
    & \nabla_{X}
    & X \ar@{|->}[l] \\
    \makebox[0pt][r]{$i\theta(\cX-\nabla_{\rho(\cX)})=$}-\alpha(\cX)
    & \cX \ar@{|->}[l]
    &       } 
\end{align}
This mean that any derivation can be decomposed in the following way:
\begin{align}
  \cX &= \nabla_{X} + ad_{\gamma} \ ,
\label{decomposition-of-derivations}
\end{align}
with $X=\rho(\cX) \in \Gamma(M)$ and $\gamma=-\alpha(\cX)\in \aA_{0}$.

We can also consider the dual exact sequence of $\cZ(\aA)$-modules:
\begin{equation*}
  \xymatrix@R=2pt{ 
    0 \ar[r]
    & \Omega^{1}(M,\End\fE)   \ar^-{\rho^{*}}[r]
    & \Oder^{1}(\aA)    \ar^-{ad^{*}}[r]  
    &\aA\otimes_{\cZ(\aA)}
    \aA^{*}_{0} \ar[r] 
    & 0\\
    &a \ar@{|->}[r]
    &a\circ \rho 
    &
    &
    \\
    &
    &\omega  \ar@{|->}[r]
    &\omega \circ \ad 
    &
  } 
\end{equation*}
where $\Omega^{1}(M,\End\fE)=\aA\otimes_{\cZ(\aA)}\Gamma(T^{*}M)$ consists of $\End(\fE)$-valued tensorial $1$-forms over $M$.
We can split this exact sequence with an ordinary connection on $\fE$ as we have done for derivations:
\begin{equation*}
  \xymatrix@R=2pt{ 
    \Omega^{1}(M,\End\fE)  
    & \Oder^{1}(\aA)    \ar_-{\nabla^{*}}[l]
    &\aA\otimes_{\cZ(\aA)} \aA^{*}_{0}\ar_-{-\alpha^{*}}[l]  
    \\
    \omega\circ\nabla 
    &\omega \ar@{|->}[l] 
    &
    \\
    &-\omega_{\Int}\circ\alpha
    &\omega_{\Int}\ar@{|->}[l]
  } 
\end{equation*}
Then we can decompose any $1$-form $\omega\in\Oder^{1}(\aA)$ in the following way:
\begin{align}
  \omega = \rho^{*}\omega^{M} - \alpha^{*}\omega_{\Int}
  \label{eq:decomposition-form}
\end{align}
with $\omega^{M}=\nabla^{*}\omega  \in  \Omega^{1}(M,\End\fE)$ and  $\omega_{\Int}=\ad^{*}\omega \in \aA\otimes_{\cZ(\aA)} \aA^{*}_{0}$.
We can remark that the space $\aA\otimes_{\cZ(\aA)} \aA^{*}_{0}$ is a space of sections of a vector bundle over $M$ with fibers $M_{n}\otimes \ksl_{n}^{*}$.
This decomposition can be easily generalize to the space $\Oder(\aA)$.

\subsubsection{Covariant differential and ordinary curvature}\label{sec:covar-diff-ordin}
The $1$-form $\alpha$ associated to any ordinary connection $\nabla$ can be fully characterized by the property: 
\begin{align}
  \alpha(ad_{\gamma}) &=-\gamma &
\forall \gamma \in \aA_{0}.
\label{eq:vertical-condition}
\end{align}
We can also characterize the space of  horizontal derivations by $(\der(\aA))_{\hor}\simeq\ker(\alpha)$.
From this properties, $\alpha$ can be compared with the connection $1$-form on a principal fiber bundle.
We will see that this noncommutative $1$-form plays here exactly the same role.
Let's first introduce the following definition:%
\begin{defn}[Covariant derivative]
  Let $\omega\in\Oder^{p}(\aA)$ and $\cX_{1}, \dots , \cX_{p+1}\in \der(\aA)$.
  The covariant differential of $\omega$ is defined by:
  \begin{equation}
\begin{aligned}
    D: \Oder^{p}(\aA) &\longrightarrow (\Oder^{p+1}(\aA))_{|\hor} \\
    \omega & \longmapsto D\omega:  (\cX_{1}, \dots , \cX_{p}) \mapsto \hd\omega(\nabla_{\rho(\cX_{1})}, \dots , \nabla_{\rho(\cX_{p+1})}) \ ,
  \end{aligned}
\end{equation}
  with $(\Oder^{p+1}(\aA))_{|\hor}$ the horizontal sub-space of $\Oder^{p+1}(\aA)$ for the action of $\Int(\aA)$.
\end{defn}

  We define the curvature noncommutative $2$-form $\Omega$ to be the covariant derivative of the noncommutative connection $1$-form $\alpha$. With the vertical property~(\ref{eq:vertical-condition}) of $\alpha$, we have:
  \begin{align}
    \Omega &=D\alpha = \hd\alpha + \alpha^2 = \rho^{*} F \ .
  \end{align}
  where $F\in \Omega^{2}_{dR}(M,\ad(P))$ is the tensorial $2$-form defined by:
\begin{align*}
  F(X,Y) &= \Omega(\nabla_{X},\nabla_{Y}) \\ 
  &=D\alpha(\nabla_{X},\nabla_{Y}) =  - \alpha([(\nabla_{X},\nabla_{Y}])
\end{align*}
for all $X, Y \in \Gamma(TM)$.
This equation can be interpreted as the obstruction to construct a Lie algebra with horizontal derivations.

Naturally, the tensorial $2$-form $F$ coincide with the tensorial $2$-form associated to the connection $\nabla$.


\subsubsection{Ordinary gauge transformations}
\label{sec:ordin-gauge-transf}
It is natural to use the algebra $\aA$ associated to the principal fiber bundle  $P$ to describe the gauge group $\cG$ and its Lie algebra $Lie(\cG)$.
Obviously, elements of $\cG$ are sections of the associated fiber bundle $P\times_{Ad} SL(n)$ which are exactly elements of $SL(\aA)$, the group\footnote{For a $SU(n)$ principal fiber bundle, the gauge group is isomorphic to $U(\aA)$, the group of unitary elements of $\aA$.}
 composed of determinant $1$ elements of $\aA$. So we have $\cG=SL(\aA)$. 
We have also that  $Lie(\cG)=\Gamma(P\times_{ad} \ksl(n))=\ksl(\aA)=\aA_{0}\simeq \Int(\aA)$.

Then, the gauge group acts infinitesimally on a connection on $P$ by Lie derivatives in the direction of internal derivations on the affine space of noncommutative $1$-forms which satisfy the vertical condition~(\ref{eq:vertical-condition}).
An element $\xi\in \aA_{0}$ act on a noncommutative $1$-form $\alpha$ which represent a connection $\nabla$ by: 
\begin{align*}
  \alpha \mapsto  \alpha^{\xi}=- \cL_{ad_\xi} \alpha = \hd \xi + [\alpha, \xi]= D\xi \ ,
\end{align*}
and  $\alpha^{\xi}$ represent the connection $\nabla^{\xi}$.

We will now introduce the concept of noncommutative tensorial forms which will permits us to generalize the action of $\cG$ on tensorial forms.

\subsection{Noncommutative tensorial forms}\label{sec:nonc-tens-forms}
\subsubsection{Associated vector bundles and representations of $\der(\aA)$}
In this section, we will see that it is possible to construct a representation of the Lie algebra $\der(\aA)$ from any vector bundle $F=P\times_{R}V$ associated to $P$ for a representation $R$.
We will consider the module $\sF=\Gamma(F)$ over $C^{\infty}(M)$ and the action of $\der(\aA)$ given by:
\begin{align}
  R(\cX)\cdot m &= \nabla^{F}_{\rho(\cX)} m - R(\alpha(\cX)) \cdot m &&
&\forall m\in \cF 
\label{eq:representation-of-derivations}
\end{align}
where $\nabla$ is a connection on $P$ and $\nabla^{F}$ its representation on sections of $F$.
This expression make sens because $\alpha$, which is the noncommutative $1$-form canonically associated to $\nabla$, takes its values in $\aA_{0}$ which naturally acts on $\cF$ by infinitesimal gauge transformations. 
This representation generalize in fact the decomposition obtained in the formula~(\ref{decomposition-of-derivations}) that we could rewrite:
\begin{align*}
  \cX = \nabla^{\End(E)}_{\rho(\cX)} - \ad_{\alpha(\cX)}
\end{align*}

On can check that the action of derivations describe in formula~(\ref{eq:representation-of-derivations})  is a representation of Lie algebra:
\begin{align*}
  R([\cX,\cY]) = [R(\cX),R(\cY)] \ .
\end{align*}
and that it is independent of the choice of the connection $\nabla$.

\subsubsection{Generalization of tensorial forms}
\label{sec:gener-tens-forms}

From the representation $R$ of $\der(\aA)$ on sections of a module $F$ associated to $P$, we can construct a differential graded complex $(\Oder(\aA,\cF), \hd^{F})$.
We define  $\Oder(\aA,\cF)$ as the set of $\cZ(\aA)$-multilinear applications from $\der(\aA)$ to $\cF$ and the differential $\hd^{F}$ by a Koszul formula: 
\begin{multline}
  \hd^{F}\omega(\cX_1, \dots , \cX_{n+1}) = \sum_{i=1}^{n+1} (-1)^{i+1} R( \cX_i) \cdot  \omega( \cX_1, \dots \omi{i} \dots, \cX_{n+1}) \\
  + \sum_{1\leq i < j \leq n+1} (-1)^{i+j} \omega( [\cX_i, \cX_j], \dots \omi{i} \dots \omi{j} \dots , \cX_{n+1}) \ .
  \label{differential-R}
\end{multline}
As usually, we can also define Cartan operations $i$ and $\cL^{R}$ on this complex.

We can define  a notion of horizontal forms and of covariant derivation on $\Oder(\aA,\cF)$, with the same definitions than in section~\ref{sec:role-ordin-conn}.
The covariant derivation that we obtain is related to the usual covariant derivative on tensorial forms by the formula:
\begin{align*}
  \rho^{*} D\omega &= \nabla^{F} \rho^{*} \omega &
  &\forall \ \omega \in \Oder(\aA,\cF)
\end{align*}

We extend the action of $Lie(\cG)$ on $\cF$ to an action on  $\Oder(\aA,\cF)$ by the formula: 
\begin{align}
\omega^{\gamma} &=  -\cL^{R}_{ad_\gamma} \omega  \ ,
\label{eq:action-Lie}
\end{align}
for $\gamma\in \aA_{0}$ and $\omega \in \Oder(\aA,\cF)$.
This action is obviously compatible with the action of $Lie(\cG)$ on tensorial form in the sens that for an element $\omega\in \Omega(M,F)$, we have:
\begin{equation}
\begin{aligned}
  (\rho^{*}\omega)^{\gamma} 
&=- \cL^{R}_{ad_\gamma} (\rho^{*}\omega)\\
&=- i_{\ad_{\gamma}} \hd^{R} (\rho^{*}\omega)
=- R(\ad_{\gamma})\cdot \rho^{*}\omega
= R(\gamma) \cdot \rho^{*}\omega \\
&= \rho^{*} \omega^{\gamma}
\end{aligned}
\end{equation}
where $\omega^{\gamma}$ denotes the usual infinitesimal gauge transformation on tensorial forms and we have used definitions  (\ref{eq:action-Lie}), (\ref{differential-R}) and (\ref{eq:representation-of-derivations}) .

\subsubsection{Decomposition of noncommutative tensorial forms}

When we have a connection $\nabla$ on $\End(E)$, we can decompose any tensorial $1$-form into vertical and horizontal part as we have done in section~\ref{sec:splitt-exact-sequ}.
Hence, an element $\omega\in \Oder(\aA,\cF)$ can be decomposed in the following way:
\begin{align}
  \omega(\cX)= a(\rho(\cX)) - \varphi(\alpha(\cX))
  \label{eq:decomposition-tensorial-ncc}
\end{align}
with $a\in \Omega^{1}(M, F)$ and  $\varphi \in \cF\otimes_{\cZ(\aA)}\cA_{0}^{*}$.
It can be written $\omega=\rho^{*}a - \alpha^{*}\varphi$ and $a$ and $\varphi$ are characterize by the relations:
    \begin{align*}
      &a = \omega \circ \nabla = \nabla^{*} \omega\\
      &\varphi = \omega \circ ad = \ad^{*} \omega
    \end{align*}
    
    With this decomposition, we can establish a relation between de differential $\hd^{F}$ and the covariant derivative on $\Oder^{1}(\aA,\cF)$.
    \begin{align}
      \hd^{F}\omega
      &= D\omega -[\alpha,\omega] - (\ad^{*}\omega) \circ \alpha^{2} + \rho^{*} \nabla (\ad^{*}\omega) \circ \alpha \\
      &= D\omega -[\alpha,\omega] - \varphi \circ \alpha^{2} + \rho^{*}( \nabla \varphi) \circ \alpha 
      \label{eq:relation-Dd}
    \end{align}
    where $ \nabla \varphi$ means the usual covariant derivative on $\varphi$ seen as a section on the vector bundles $F\otimes \End_{0}^{*}(E)$ ($\End_{0}^{*}(E)$ is the dual bundle of the bundle of traceless endomorphisms).
This relation will be useful in section~\ref{sec:nonc-conn} for the calculation of the noncommutative curvature of a noncommutative connection.

\begin{rem}
From the definition given in \ref{sec:endom-algebr-definition}, we can
 associate to every associated vector bundle $F$ an endomorphism algebra
 $\aA^{F}$ which is a $\der(\aA)$ module and we can consider the differential graded algebra $\Oder(\aA,\aA^{F})$ as a differential calculus for the algebra $\aA^{F}$.
  
\end{rem}
  
\subsection{Chern-Weil homomorphism}\label{weil-homomorphism}
We have seen that the exact sequence of derivations (\ref{der-exact-sequence}) can always be split as a sequence of $\cZ(\aA)$-modules with an ordinary connection but that there is an obstruction to spilt it as an exact sequence of Lie algebras. 
This obstruction is characterized by the curvature of this connection.
We will show in this section that this exact sequence can be used to construct a Chern-Weil homomorphism using the construction presented in \cite{lecomte:85} adapted to the short exact sequence:
\begin{equation}
  \xymatrix@1{ {0} \ar[r] & \aA_{0} \ar[r]^-{ad} & \der(\aA) \ar[r]^-{\rho} & \Gamma(TM) \ar[r] & {0}} \ ,
  \label{eq:shortseqchernweil}
\end{equation}


First, we can remark that the Lie algebra $\aA_{0}$ is a $\der(\aA)$-module and a $Z(\aA)$-module.
Thus we can consider the complex of $Z(\aA)$ multilinear symmetric applications from $\aA_{0}$ to $Z(\aA)$ which we denote:
\begin{align*}
\cC^{q}=  S_{Z(\aA)}^{q} \aA_{0}^{\star Z(\aA)} \ ,
\end{align*}
where $\star Z(\aA)$ is the operation which acts on the category of $Z(\aA)$-modules described in
 \cite{dubois-violette:japan99}.
We have a natural action of $\der(\aA)$ on this complex which is given by a Lie derivative.
So it is natural to consider the invariant sub-complex $(S_{Z(\aA)}^{q} \aA_{0}^{\star Z(\aA)})_{\der(\aA)-\inv} $.
An element $f$ of this complex, is a $Z(\aA)$ multilinear symmetric applications from $\aA_{0}$ to $Z(\aA)$ which is invariant under the action of $\der(\aA)$:
\begin{align}
  L_{\cX}f &= 0&
  & \forall \ \cX \in \der(\aA)  \ ,
\label{eq:invcond}
\end{align}
where $L_{\cX}$ is the Lie derivative naturally defined on multilinear applications.
This space is isomorphic to the space of invariant polynomials on $\ksl(n)$.
To show that, we can introduce an ordinary connection $\nabla$ which split the short exact sequence~(\ref{eq:shortseqchernweil}).
then the invariant condition (\ref{eq:invcond}) can be split into a vertical and horizontal part and we have:
\begin{align*}
 &\left\{ \begin{aligned}
    L_{\nabla_{X}}f &= 0 \\
    L_{\ad_{\gamma}}f &=0
\end{aligned}\right. &
&\forall X\in \Gamma(M), \gamma \in \aA_{0} \ .
\end{align*}
We deduce from this relations that $f$ is a constant application on the base manifold $M$ and that it define an invariant polynomial on $\ksl(n)$.

Then, the curvature $2$-form $F$ associated to $\nabla$ permit to associate to $f$ an element: 
\begin{align*}
  f_{\nabla}= f(F \wedge \dots \wedge F)\in   \Omega^{2q}(M) \ , 
\end{align*}
We then recover the usual construction of the Chern-Weil homomorphism and so we have that $f_{\nabla}$ is a closed form and that its cohomology class is  independent of  $\nabla$.
In conclusion, we have construct a linear application: 
\begin{align*}
  P_{I}(\ksl(n)) \simeq (S_{Z(\aA)}\aA_{0}^{\star Z(\aA)})_{\der(\aA)-\inv} &\longrightarrow   H^{\text{even}}(M)\\
  f &\longmapsto [f_{\nabla}]
\end{align*}
which associate to every invariant polynomial on $\ksl(n)$ an element of the even de~Rham cohomology of $M$.
Obviously, this application correspond to usual Chern-Weil homomorphism. 

We can notice that usually, this homomorphism can be obtained from the Weil algebra which use the finite dimensional Lie algebra the structure group.
Here we didn't use this Lie algebra but we have worked directly with the Lie algebra $\Int(\aA)\simeq \aA_{0}$ which can be identify with the Lie algebra of infinitesimal gauge transformations.
This Lie algebra is infinite dimensional and we have surprisingly extract from it invariant polynomials by imposing an invariance condition with respect to the (infinite dimensional) Lie algebra  $\der(\aA)$.

This construction raise new questions.
For an endomorphism algebra $\aA$, it could be interesting to see if this construction could be related to other homological constructions for algebras like the basic cohomology introduced in \cite{dubois-violette:94}, where it is shown that basic cohomology of the algebra $M_{n}(\gC)$ is isomorphic to the set of invariant polynomials on $M_{n}(\gC)$.
There is also an other approach \cite{crain:98} of the Chern-Weil theory in noncommutative geometry which use Hopf algebra actions and cyclic cohomology. This context is more general than the one considered here in the sens that it can be adapted to foliations or to pure algebras with an Hopf algebra action. 
The algebra which was considered in \cite{crain:98} to construct the Chern-Weil map is the convolution algebra of the holonomy groupoid of a foliation. In the case of a principal fiber bundle, this algebra is Morita equivalent to the space of functions on the base manifold.
It could be interesting to make a contact with this approach of the Chern-Weil theory and the one presented here.
This could be done by taking into account the Lie algebroid structure of $\der(\aA)$ and its relation with the Atiyah Lie algebroid remarked in \cite{dubois-violette:98}.

We can also remark that this construction can be easily transposed to any associative algebra $\aA$, using the short exact sequence of derivations:
\begin{equation*}
\xymatrix@1{ {0} \ar[r] & \Int(\aA) \ar[r] & \der(\aA) \ar[r]^-{\rho} & \Out(\aA)  \ar[r] & {0}} \ .
\end{equation*}
This would lead to interesting results if the space of internal derivations is not too small.

 \section{Noncommutative connections}\label{sec:nonc-conn}

In the previous section, we have introduce a notion of ``ordinary connection'' where the notion of connection referred to splittings of short exact sequences of Lie algebras and $\cZ(\aA)$-modules.
We will now introduce an other notion of connection for endomorphism algebras.
 
For a right $\aA$-module $\cM$, we consider the following definition of connection:
\begin{defn}
  We call a $\Oder(\aA)$-connection, an application:
\begin{align*}
  \hat\nabla: \cM &\to \cM \otimes_{\aA}\Oder^{1}(\aA)
\end{align*}
which satisfies:
\begin{align*}
  \nabla(m b)&=\nabla(m)b +m \hd b \ .
\end{align*}
\end{defn}

Such a connection can naturally be extended to an application:
\begin{align*}
  \hat\nabla: \Oder^{n}(\aA,\cM) \longrightarrow \Oder^{n+1}(\aA,\cM)
\end{align*}
 using a Koszul formula:
\begin{align}
  \hat\nabla\omega(\cX_1, \dots , \cX_{n+1}) = \sum_{i=1}^{n+1} (-1)^{i+1} \nabla_{\cX_i} \omega( \cX_1, \dots \omi{i} \dots, \cX_{n+1}) \\
  + \sum_{1\leq i < j \leq n+1} (-1)^{i+j} \omega( [\cX_i, \cX_j], \dots \omi{i} \dots \omi{j} \dots , \cX_{n+1}) \ .
  \label{eq:Koszul-nc}
\end{align}

\subsection{Reference connection}
For an endomorphism algebra $\aA$, we can deduce from Morita equivalence that any finite right $\cA$ projective module can be put through the form $\cM=\Gamma(F\otimes E^{*})$ where $F$ is an arbitrary vector bundle over the base manifold $M$.
Locally one has $\cM \simeq C^{\infty}(M)\otimes M_{k,n}(\gC)$, where $M_{k,n}(\gC)$ are $k\times n$ complex matrices.

We can remark that any ordinary connection $\nabla^{F\otimes E^{*}}$ on $F\otimes E^{*}$ can be used to define a reference connection:
\begin{align*}
  \tilde{\nabla}_{\cX} m &= \nabla^{F\otimes E^{*}}_{\rho(\cX)} m + m \cdot \alpha(\cX).
\end{align*}
where $\alpha$ is the noncommutative $1$-form associate to the connection $\nabla^{E^{*}}$ on $E^{*}$.

\subsection{Associated modules}
We can specialize to the case where the vector bundle $F$ is associated to the principal fiber bundle $P$.
We will then say that the module $\cF=\Gamma(F)$ is associate to $P$.
 Specialization to such modules will be necessary in order to have a natural action of $\der(\aA)$ onto elements of $\aA$-module $\cM$. 
 We will write $F=P\times_{R^{F}}V$ where $R^{F}$ is a representation of $SL_{n}$ on the vector space $V$.
Then, using the fact that the space of connections is an affine space, we can write any noncommutative connection:
\begin{align}
  \hat\nabla_{\cX}m  &= \tilde{\nabla} m + B(\cX) \cdot m  .
\label{eq:ncc-B}
\end{align}
with  $B \in \Oder(\aA,\aA^{F})$ and $\aA^{F}\simeq \End_{\aA}(\cM)\simeq \Gamma(\End F)$ is the endomorphism algebra associated to the vector bundle $F$.

Because $\cM$ is now a $\der(\aA)$-module, we can introduce the differential module $(\Oder(\aA,\cM),\hd)$ and  also write a noncommutative connection:
\begin{align}
  \hat\nabla_{\cX}m 
  &= \hd m + \omega(\cX)\cdot m 
  \label{eq:ncc-omega}
\end{align}
with $\omega \in \Oder(\aA,\aA^{F})$.
We can remark that, using the formula (\ref{eq:representation-of-derivations}), the differential $\hd$ is here defined by:
\begin{align*}
  \hd m (\cX) &=  \nabla^{F\otimes E^{*}}_{\rho(\cX)} m -R^{F}(\alpha(\cX)) m + m \cdot \alpha(\cX)  \ .
\end{align*}
Then $\omega$ and $B$ are relate by the formula:
\begin{align*}
  \omega &= B + R^{F}(\alpha)
\end{align*}
Hence, we have two possible decompositions of a noncommutative connections.
We will see that the decomposition~(\ref{eq:ncc-omega}) is well adapted to do algebraic computations.
It will be more convenient to use the decomposition~(\ref{eq:ncc-B}) when we will write a Yang-Mills action (cf. section~\ref{sec:noncommutative-yang}) for noncommutative connections.
Indeed, with an ordinary connection $\nabla$ on $P$, the decomposition $B =\rho^{*}a - \alpha^{*}\varphi$ with $a\in \Omega^{1}(M,\End F)$  a tensorial $1$-form and $\varphi \in \cA^{R}\otimes \cA_{0}^{*}$ a section of a vector bundle over $M$, will permit us to distinguish Yang-Mills fields and scalar fields in a noncommutative connection. 

It is also interesting to describe the gauge group associate to a noncommutative connection.
The group $\Aut(\cM)$ of automorphism of $\cM$ acts on the affine space of connections:
\begin{align*}
  \widehat{\nabla} \mapsto \widehat{\nabla}^U = U^{-1}\circ \widehat{\nabla} \circ U \ .
\end{align*}
Because $\cM$ has an hermitian structure, it is natural to restrict us to the action of unitaries of $\Aut(\cM)$.
So, we will call the ``noncommutative gauge group''  $\hat\cG=U(\cA^{R})$.
The action of $\hat\cG$ can be transposed to the noncommutative $1$-form $\omega$ defined in formula (\ref{eq:ncc-omega}).
The $1$-form $\omega^{U}$ which represent $ \widehat{\nabla}^U$ is related to $\omega$ by the formula:
\begin{align}
  \omega^{U} = U^{-1} \omega U + U^{-1} \hd U
\end{align}
One can also consider infinitesimal transformations:
\begin{align}
  \omega &\mapsto \hd \gamma+ [\omega, \gamma] \ , &
  & \text{for } \gamma \in \ksl(\aA^{R}) \ .
\label{eq:ncgaugetrans}
\end{align}

It is important to notice that elements of $\hat\cG$ are sections of a bundle associated to $P$ (with structure group $SL_{n}$ or $SU(n)$) and with fibers $U(k)$.
Hence, this group has nothing to do with the ``geometric gauge group'' $\cG=SL(\aA)$ introduced in section \ref{sec:gener-tens-forms} which is associated to the principal fiber bundle $P$. 
We have see in section  \ref{sec:ordin-gauge-transf} that this group acts naturally on noncommutative tensorial forms in a way that its action extend the action on ordinary tensorial forms. 
Its infinitesimal action corresponds to Lie derivatives in the direction of internal derivations, as in formula~(\ref{eq:action-Lie}). 
So, for an element $\omega\in \Oder^{1}(\aA,\cM)$, it corresponds to the action:
\begin{align}
  \omega &\mapsto  -\cL^{R}_{ad_\gamma} \omega  &
  & \text{for } \gamma \in \ksl(\aA) \ .
  \label{eq:ordgaugetrans}
\end{align}

We can remark that the action (\ref{eq:ordgaugetrans}) and (\ref{eq:ncgaugetrans}) coincide when $\aA^{R}=\aA$ and when $\omega$ is a noncommutative $1$-form which represent  an ordinary connection, so which satisfies the vertical condition $\omega(\ad_{\gamma})=- \gamma$.
We can also remark that the action~(\ref{eq:ordgaugetrans}), correspond to what we usually call ``active'' gauge transformation on tensorial objects and that it has a ``passive'' counterpart which is the fact that the transformation (\ref{eq:ordgaugetrans}) have local expressions similar to changes of trivializing charts.

\subsection{Noncommutative curvature}
    
    In the case of a $P$ associated module, we can characterize the curvature by a noncommutative $2$-tensorial form $\Omega$ by posing:
    \begin{align}
      (\hat{\nabla}^{2} m) (\cX,\cY) = \left( [\hat{\nabla}_{\cX},\nabla_{\cY}-\hat\nabla_{[\cX,\cY]}\right) m = \Omega(\cX,\cY) \cdot m
      \label{eq:nc-curvature}
    \end{align}
    with $\Omega\in \Oder^{2}(\aA,\aA^{F})$.
    Then if we use the decomposition 
    \begin{align*}
      B=\rho^{*}a - \varphi \circ \alpha
\end{align*}
    associated to a connection $\nabla$, a direct calculation show that:
    \begin{multline}
      \Omega = \rho^{*}\biggl( R(F) - \varphi(F) + \nabla a + a^{2}\biggl)- \alpha^{*}\rho^{*}\biggl(\nabla \varphi+[a,\varphi]\biggl)  +\\+\frac{1}{2} \biggl([\varphi\circ\alpha,\varphi\circ \alpha] - \varphi\circ[\alpha,\alpha]\biggl) \ .
      \label{eq:nc-curvature-calc}
    \end{multline}
    where $F$ is the curvature tensorial $2$-form associated $\nabla$.
    
    We can also calculate the curvature $\Omega$ with the N.C. tensorial form $\omega$ defined in equation~(\ref{eq:ncc-omega}), and so, from (\ref{eq:nc-curvature}), we have:
    \begin{align*}
      \Omega &= \hd \omega + \omega^{2}
    \end{align*}
    where $\hd$ is here the differential defined on $\Oder(\aA,\aA^{F})$.
    Now, if we the decomposition associate to a connection $\nabla$: 
    \begin{align*}
      \omega= R^{F}(\alpha)+\rho^{*}a -\varphi\circ\alpha
    \end{align*}
    we obtain that:
    \begin{align*}
      \Omega &= \hd \omega + \omega^{2} \\
      &= D\omega -[\alpha,\omega] + (1-\varphi)\circ \alpha^{2} - \rho^{*}\nabla\varphi\circ \alpha+
      +[\alpha,\omega] - \alpha^{2} + \rho^{*}a^{2}  + (\varphi\circ\alpha)^{2} - [\rho^{*}a,\varphi\circ\alpha]) \\
      &= D\omega +\rho^{*}a^{2} - \left( \rho^{*}\nabla\varphi\circ \alpha +  [\rho^{*}a,\varphi\circ\alpha]  \right) +  (\varphi\circ\alpha)^{2} - \varphi\circ \alpha^{2} 
    \end{align*}
    This result is the same than in the formula (\ref{eq:nc-curvature-calc}), thanks to the relations:
    \begin{align*}
      & D R^{F}(\alpha) = \rho^{*} R^{F}(F) \\
      & D \rho^{*}a = \rho^{*}\nabla a  \\
      & D (\varphi\circ \alpha) = \varphi\circ R^{F}(F)
    \end{align*}
    
    It is interesting to give local expressions of the curvature.
    Locally, in the basis of derivations $\{\nabla_{\mu},ad_{E_{a}}\}$ where
    $\nabla_{\mu}=\nabla_{\partial_{\mu}}$ and $\{E_{a}\}$ is an hermitian basis of $\ksl_{n}(\gC)$, on has:
    \begin{equation*}
    \begin{aligned}
      \Omega_{\mu \nu} &=\Omega(\nabla_{\mu}, \nabla_{\nu})&
      & = R(F_{\mu \nu}) - \varphi(F_{\mu \nu}) + \nabla_{\mu}a_{\nu}- \nabla_{\nu}a_{\mu}+[a_{\mu}, a_{\nu}]  \\
      \Omega_{\mu b} &= \Omega(\nabla_{\mu}, \ad_{E_{b}})&
      & = \nabla_{\mu}\varphi_{b} + [a_{\mu}, \varphi_{b}] \\
      \Omega_{a b} &=  \Omega(\ad_{E_{a}}, \ad_{E_{b}}) &
    &= [\varphi_{a},\varphi_{b}] -\varphi_{c}C_{ab}^{c} 
  \end{aligned}
\end{equation*}
where:
\begin{equation*}
  \begin{aligned}
    \varphi_{a}&=\varphi(E_{a})\ , &
    \nabla_{\mu}\varphi_{b}&= \partial_{\mu}\varphi_{b}+[R(A_{\mu}),\varphi_{b}]- C_{ab}^{c}A_{\mu}^{a}\varphi_{c}\\
    a_{\mu} &= a(\partial_{\mu})\ , &
    \nabla_{\mu}a_{\nu} &= \partial_{\mu}a_{\nu} + [R(A_{\mu}),a_{\nu}]
  \end{aligned}
\end{equation*}
and $A$ is the local gauge potential associated to $\nabla$.

Now, we would be interested into write a Yang-Mills-like action for this noncommutative connection.
In order to do that, we need before to introduce a notion of metric and of integration.

\section{Integration and Riemannian structure}
\label{sec:integr-riem-struct}
\subsection{Riemannian structure}\label{sec:riemannian-structure}
It is possible to introduce a notion of metric on $\der(\aA)$ and on $\Oder(\aA)$ which generalize the notion of metric on vector fields on $M$ and on de~Rham differential forms.
We will need this notion to be able to construct a Yang-Mills type action for noncommutative connections.

We will see that this notion of metric will produce a mechanism similar to the one encounter in Kaluza-Klein theories over principal fiber bundles.

\subsubsection{Metrics}
\begin{defn}
  We will call a (pseudo-)metric on $\der(\aA)$, a  symmetric $Z(\aA)$-bilinear application (\textit{i.e.} an element of $(S^{2}_{Z(\aA)}\der(\aA))^{\star Z(\aA)}$ ):
  \begin{align*}
    g: \der(\aA)\otimes_{Z(\aA)}\der(\aA) \longrightarrow Z(\aA)
  \end{align*}
  We will say that this metric is non degenerate if:
  \begin{align*}
    g^{\flat}:  \der(\aA)&\longrightarrow \Oder^{1}(\aA) \\
    \cX &\longmapsto [\cY \mapsto g(\cX,\cY)]  
  \end{align*}
  is injective and $\Oder^{1}(\aA)$ is span by its $Z(A)$-submodule $(\Im g^{\flat})$.
\end{defn}

It is known that $\der(\aA) \simeq (\Oder^{1}(\aA))^{\ast \aA}=\Hom_{\aA}^{\aA}(\Oder^{1}(\aA),\aA)$ and then, $\der(\aA)$ can be consider as a sub module of $\Hom_{Z(\aA)}^{Z(\aA)}(\Oder^{1}(\aA),\aA)$.
In this sense, we can see $\der(\aA)$ as a $\aA-\aA$ bimodule, which is here isomorphic to $\aA\otimes_{\cZ(\aA)}\der(\aA)$.
We can easily extend $g$ to an homomorphism of $\aA$-bimodule\footnote{In order to simplify the notation, will use the notation $\der(\aA)$ for the $\aA$-bimodule generate by $\der(\aA)$.}:
\begin{align*}
  g: \der(\aA)\otimes_{\aA}\der(\aA) \longrightarrow \aA \ ,
\end{align*}
by imposing $g(a\cdot \cX \cdot b, c \cdot \cY \cdot d)= a \cdot b \cdot c \cdot d \cdot g(\cX,\cY)$ for all $a,b,c,d \in \aA$ and all $\cX,\cY \in \der(\aA)$, with $g(\cX,\cY)\in \cZ(\aA)$.
Then the non degeneracy condition on $g$ is equivalent to say that $g^{\flat}$ is a $\aA$ bimodule homomorphism which is bijective.
We will call $h^{\#}$ its inverse.
Then we can define a notion of metric on $\Oder^{1}(\aA)$ by considering the $\aA$-bimodule homomorphism:
\newcommand{\ho}{h}
\begin{align*}
  \ho:  \Oder^{1}(\aA)\otimes_{\aA}\Oder^{1}(\aA) \longrightarrow \aA\\
  (\omega,\eta) \longmapsto  g(\ho^{\#}(\omega),\ho^{\#}(\eta))
\end{align*}
This notion of metric can easily be extend to a notion of metric on $\Oder(\aA)$ as an homomorphism of $\aA$-bimodule:
\begin{align*}
  \ho:  \Oder(\aA)\otimes_{\aA}\Oder(\aA) \longrightarrow \aA
\end{align*}
For that, we pose for elements of $\Oder(\aA)$ of the form $\omega = \Pi_{i} \omega^{(i)}$ with $\omega^{(i)}\in \Im(g^{\flat}) \subset \Oder^{1}(\aA)$:
\begin{equation*}
  \ho(\omega,\eta)= 
  \left\{
     \begin{aligned}
       &O & 
       &\text{if $deg(\omega)\neq deg(\eta)$}\\
       &\det((\ho(\omega^{(i)},\eta^{(j)}))_{i,j\in [1,N]} ) & 
       &\text{for $n=deg(\omega)= deg(\eta)$}
     \end{aligned}\right.
 \end{equation*}
We then extend this definition to all elements of $\Oder(\aA)$ by linearity and action of $\aA$.

We will say that $h$ is the metric inverse of $g$ and that $g$ and $h$ define on $\aA$ a (pseudo-) Riemannian structure.

\subsubsection{Decomposition of the metric and reference connection}\label{sec:decomp-metr-refer-1}

We will show the following proposition:

\begin{prop}\label{prop:decomp-metr-refer}
  For a Riemannian structure on $\aA$ given by a metric $g$ on $\der(\aA)$, such that $g_{\Int}=ad^{*}g$ is non degenerate, then there exist a unique connection $\nabla$ on $\End(\fE)$ such that:
  \begin{align}
    g(\nabla_{X}, \ad_{\gamma})&= 0 &
    & \forall X\in \Gamma(M) ,  \gamma \in \aA 
    \label{eq:orthogonalite} 
  \end{align}
\end{prop}
\demonstration{\noindent {\slshape Demonstration : }}
\demo

The application $g_{\Int}$ is a $\cZ(A)$-bilinear application from $\aA_{0}$ to $\cZ(\aA)$ and we can associate to it an application:
$g^{\flat}_{\Int}: \aA_{0}\to  \aA_{0}^{\star\cZ(\aA)}$.
We will denote $h^{\#}_{\Int}$ its inverse.
As we have done for $h^{\#}$, we can extend $h_{\Int}^{\#}$ to an homomorphism $\aA\otimes_{\cZ(\aA)} \aA_{0}^{\star}\to \aA\otimes_{\cZ(\aA)}\aA_{0}$ and define the inverse metric $h_{\Int}$ of $g_{\Int}$.
Then it make sens to consider the application:
\begin{align*}
 \alpha(\cX)
= - g(\cX, \ad \circ h^{\#}_{\Int}(\Id_{\aA_{0}})) 
= - h_{\Int}(ad^{*}g^{\flat}(\cX),\Id_{\aA_{0}})
\end{align*}
where the element $ad \circ h^{\#}_{\Int}(\Id_{\aA_{0}})$ is in the $\aA$ bimodule generate by $\Int(\aA)$.
We can check by a simple calculation that $\alpha(\ad_{\gamma})=-\gamma$ for all $\gamma \in  \aA_{0}$.
Then $\alpha$ define a connection $\nabla$ on $\End(\fE)$ which is given by the formula:
\begin{align*}
  \nabla_{\rho(\cX)}&= \cX + \ad(\alpha(\cX)) &
& \forall \cX\in\der(\aA) \ .
\end{align*}
This connection satisfies the relation (\ref{eq:orthogonalite}) and the non degeneracy of $g_{\Int}$ shows its uniqueness.
\findem

From the previous proposition, we deduce the following property on every metric on $\der(\aA)$ which is non degenerate along fibers, \textit{i.e.} for which the metric $\ad^{*}g$ is non degenerate.
\begin{prpt}
  If $g$ is a metric non degenerate along fibers, then it can be decomposed as:
\begin{align*}
  g &= \rho^{*} g^{M} + \alpha^{*} g_{\Int} \ ,& 
  &\text{with} &
  & g^{M}=\nabla^{*}g &
  & \text{and} &
& g^{\Int}=\ad^{*}g \ .
\end{align*}
\end{prpt}
We can remark that this decomposition is similar to the one previously encounter for noncommutative $1$-forms.

We also have the following property:
\begin{prpt}  
  For a metric $g$ as in proposition~\ref{prop:decomp-metr-refer} and with inverse metric $h$, we have that:
  \begin{align*}
    &g^{M \flat}=\nabla^{*}\circ g^{\flat}\circ\nabla &
    &\text{has for inverse} &
    & h^{M\#}=\rho\circ h^{\#}\circ\rho^{*} \\
    &h_{\Int}^{\#}=\alpha\circ h^{\#}\circ\alpha^{*} &
    &\text{has for inverse} &
    &g_{\Int}^{\flat}=\ad^{*}\circ g^{\flat}\circ ad
  \end{align*}
  The  equation~(\ref{eq:orthogonalite}) has its equivalent for $h$, which is:
  \begin{align*}
    \ho(\rho^{*}\mu,\alpha^{*}\eta)&=0 &
    &\forall \mu \in \Omega(M,\End(\fE)), \forall \eta\in \aA\otimes_{\cZ(\aA)}\aA_{0}^{*} \ .
    \label{eq:orthogonalite:2}
  \end{align*}
  Finally, we have that the metric $h$ on $\Oder^{1}(\aA)$ can be decomposed in the following way:
  \begin{align*}
    h &= \nabla_{*} h^{M} + \ad_{*}h_{\Int} \ , & 
    &\text{with} &
    & h^{M}=\rho_{*}h &
    & \text{and} &
    & h_{\Int}=\alpha_{*}h \ .
  \end{align*}
\end{prpt}

\begin{rem}
We can remark that it is possible to construct a metric from an ordinary metric $g^{M}$ on $M$ and a connection $\nabla$ on $P$ and a non degenerate symmetric bilinear form on $\aA_{0}$.
\end{rem}

\begin{rem}
  The Killing form on $\ksl_{n}$ define a particular internal metric which has the particularity to being invariant through the action of $\der(\aA)$.
  We can compare this internal metric as an equivalent of the internal $1$-form $i\theta$ which is also $\der(\aA)$-invariant.
\end{rem}
\subsubsection{Local expressions of the metric}
\label{sec:local-expr-metr}
We will give local expressions of a metric satisfying the conditions of proposition~\ref{prop:decomp-metr-refer}.
We will work over an open chart $U$ with coordinates ${x^{\mu}}$ and a basis ${E_{a}}$ of hermitian matrices of $\ksl_{n}$.
In this basis, we can describe the metric $g\loc$ by its components:
\begin{align*}
  g\loc(\partial_{\mu},\partial_{\nu}) &= g_{\mu  \nu} &
  g\loc(\partial_{\mu},\ad_{E_{b}}) &=g_{\mu b} \\
  g\loc(\ad_{E_{a}},\partial_{\nu}) &= g_{a \nu} &
  g\loc(\ad_{E_{a}},\ad_{E_{b}}) &= g_{a b}
\end{align*}
If we denote $\cB=(\partial_{\mu},\ad_{E_{a}})$ the local basis of derivations, we will denote the components of the metric by:
\begin{align*}
  &(g\loc)_{\cB} = \left(  \begin{matrix}
      g_{\mu  \nu} & g_{\mu b} \\
      g_{a \nu} & g_{a b}
    \end{matrix}\right )&
\end{align*}

Now, from the previous proposition, we can define a noncommutative $1$-form $\alpha$ which will have for local expression:
\begin{align*}
  \alpha\loc &= E_{a} \alpha^{a}  = E_{a} (A^{a} - i\theta^{a})
\end{align*}
We denote $\cB^{\star}=(dx^{\mu}, i\theta^{a})$ the dual basis of $\cB$, and $A^{a}$ is a local $1$-form in $\Omega(U)\otimes \ksl_{n}$, called the gauge potential.
It is defined by:
\begin{align*}
  A^{a}_{\mu} &= -h_{\Int}^{ab}g_{b \mu}  &
  &\text{and} &
& (h_{\Int}^{ab}) = (g_{ab})^{-1} \ .
\end{align*}

A more appropriate basis to describe the metric is obviously the basis  $\cB'=(\nabla_{\mu},\ad_{E_{a}})$, with  $\nabla_{\mu}=\nabla_{\partial_{\mu}}=\partial_{\mu}+\ad_{A_{\mu}}$, for derivations, and the dual basis  $\cB'^{\star}=(dx^{\mu},-\alpha^{a})$ for $1$-forms.
We have in this basis:
\begin{align*}
  &  (g\loc)_{\cB'} = \left(  \begin{matrix}
      g^{M}_{\mu  \nu} & 0 \\
      0 & g_{a b}
    \end{matrix}\right ) \ , &
  & \text{with}&
  &\begin{aligned}
    g^{M}_{\mu \nu} &= g\loc(\nabla_{\mu},\nabla_{\nu}) = g_{\mu \nu} - A_{\mu}^{a} A_{\nu}^{b} g_{ab}
  \end{aligned} \ \ .
\end{align*}
We denote the components of $h\loc$ in the dual basis $\cB^{\star}=(dx^{\mu},i \theta^{a})$:
\begin{align*}
    h\loc(dx^{\mu},dx^{\nu}) &= h^{\mu  \nu} &
  h\loc(dx^{\mu},i\theta^{b}) &=h^{\mu b} \\
  h\loc(i\theta^{a},dx^{\nu}) &= h^{a \nu} &
  h\loc(i\theta^{a},i\theta^{b}) &= h^{a b}
\end{align*}
We have the property that:
\begin{align*}
  A^{a}_{\mu} &= g^{M}_{\mu\nu} h^{a\nu} &
  &\text{and} &
& (g^{M}_{\mu\nu}) = (h^{\mu  \nu})^{-1} \ , 
\end{align*}
and in the dual basis  $\cB'^{\star}=(dx^{\mu},-\alpha^{a})$, $h\loc$ take the diagonal form:
\begin{align*}
  &  (h\loc)_{\cB'^{\star}} = 
  \left(  \begin{matrix}
      h^{\mu  \nu} & 0 \\
      0 & h^{a b}_{\Int}
    \end{matrix}\right ) &
  &\text{with} &
  h^{a b}_{\Int} &= h^{ab} - h^{\mu \nu} A_{\mu}^{a} A_{\nu}^{b}
\end{align*}

If we return to the basis $\cB$ and $\cB'$, the metric $g$ and $h$ have the following form:
\begin{align*}
  &(g\loc)_{\cB} = 
  \left(  \begin{matrix}
      g_{\mu  \nu} & -A_{\mu}^{a}g_{a b} \\
      -g_{a b} A_{\nu}^{b} & g_{a b}
    \end{matrix}\right )&
  &(h\loc)_{\cB} = \left(  \begin{matrix}
      h^{\mu  \nu} & h^{\mu \nu}A_{\nu}^{b} \\
      A_{\mu}^{a} h^{\mu \nu} & h^{a b}
    \end{matrix}\right ) \ .
\end{align*}

\begin{rem}
  The expressions that have obtain are similar to expressions that we could have obtain by the formulation of a Kaluza-Klein theory on the principal fiber bundle $P$.
\end{rem}

\subsection{Integration}
It is possible to define(see~\cite{masson:99}) an integration  ``along fibers'':
\begin{align*}
  \int_{nc} : \Oder(\aA) \longrightarrow \Omega(M) 
\end{align*}
In~\cite{masson:99}, the metric used to define this integration was the Killing metric.
The construction was based on the cycle, in the sens of \cite{Conn:85}, introduced in \cite{dubois-violette:90} for matrix algebras.
For an algebra of endomorphisms with an arbitrary Riemannian structure, we can generalize this construction by considering a cycle on ``fibers'' which is not necessary invariant.
Then it easy to construct a cycle $(\Oder(\aA),\fint)$ on $\aA$ by considering the application: 
\begin{align*}
  \fint : \Oder(\aA) &\longrightarrow \gC \\
  \omega &  \mapsto  \int_{M}\circ \int_{nc} \omega
\end{align*}
with $\int_{M}$ the usual integration on the manifold $M$ and $ \int_{nc}$ the integration along fibers.
One can construct this cycle in a similar way than in~\cite{masson:99} using the metric $g_{ab}=g\loc(\ad_{E_{a}},\ad_{E_{a}})$ (in local coordinates) instead of the Killing metric on matrices.


\subsubsection{Hodge operation}

As in the commutative case, when $\aA$ comes with a Riemannian structure, it is then natural to define a Hodge operation:
\begin{align}
  \star : \Oder^{k}(\aA) \to \Oder^{d+n^{2}-1 -k}(\aA) \ ,
\end{align}
with $d$ the dimension of the manifold $M$ and $n$ the rank of the fibers of $\fE$.

Locally, an element $\omega \in \Oder^{r}(\aA)$ can be write $\omega\loc = \omega_{\mu_{I},r_{J}} dx^{\mu_{I}} \alpha^{r_{J}}$, with $I$ and $J$ multi-indices.
Then the Hodge operation is define locally by the formula:
\begin{align*}
  (\star \omega)\loc =\frac{(-1)^{|J|(d-|I|)}}{(d-|I|)!(n^{2}-1-|J|)!}  \ \omega_{\mu_{I},r_{J}}  \
  h^{\mu_{I}\nu_{I}}
  \epsilon^{M}_{\nu_{I} \nu_{K}} \
  h_{\Int}^{r_{J} s_{J}}
  \epsilon_{s_{J}  s_{L}}  \
  dx^{\nu_{K}}
  \alpha^{s_{L}} \ ,
\end{align*}
with $\epsilon^{M}$ and $\epsilon$ the tensors:
\begin{align*}
  \epsilon^{M}_{\nu_{1} \cdots \nu_{d}} &= \sqrt{\det(g^{M}_{\mu \nu})} \ \delta_{\nu_{1} \cdots \nu_{d}}^{1 \cdots \  d} \\
  \epsilon_{s_{1} \cdots s_{n^{2}-1}}&= \sqrt{\det(g_{ab})} \ \delta_{s_{1} \cdots s_{n^{2}-1}}^{1 \cdots \ n^{2}-1}    \ .
\end{align*}
where the symbol $\delta_{A_{1} \cdots A_{N}}^{1 \ \cdots \ N}$ is the totally anti-symmetric Krönecker symbol (the determinant of the matrix $(\delta^{I}_{A_{J}})_{I,J\in [1,N]}$, with $\delta$ the usual Krönecker symbol).

One check easily that this form is of degree $(d+n^{2}-1-r)$ and that the Hodge operation satisfies:
\begin{align*}
  \star \star &= (-1)^{r(d+n^{2}-1-r)}
\end{align*}
on forms of degree $r$.

Now, we can write the metric in the following way:
\begin{align*}
  h(\omega, \eta) &= \star^{-1} ( \omega \star \eta) &
  &\text{if $\omega$ and $\eta$ have the same degree} \\
  &= 0 &
  &\text{elsewhere}
\end{align*}
with  $\omega, \eta \in \Oder(\aA)$.

One can also easily extend the Hodge operation to $\Oder(\aA,\cF)$, the differential complex construct for any $P$-associate module $\cF$ defined in section~\ref{sec:nonc-tens-forms}.

\subsubsection{Hermitian structure and scalar product on N.C. tensorial forms.}
Let $\cF$ a right $\aA$-module.
We call an hermitian structure on a $\cF$ an application:
\begin{align*}
  \langle \ , \ \rangle : \cF \otimes_{C^{\infty}(M)}\cF \longrightarrow \aA 
\end{align*}
We can obviously extend it on the complex $\Oder(\aA,\cF)$ to an application:
\begin{align*}
  \langle\ , \ \rangle: \Oder(\aA, \cF) \otimes_{C^{\infty}(M)} \Oder(\aA,\cF) \longrightarrow \Oder(\aA)
\end{align*}

With such an hermitian structure and a hodge operation on $\Oder(\aA,\cF)$,  we can define a scalar product on $\Oder(\aA,\cF)$ which is defined for two elements $\omega, \eta \in \Oder(\aA,\cF)$ by:
\begin{align*}
  (\omega,\eta) &= \fint \langle\omega,\star\eta\rangle
\end{align*}

\section{The noncommutative Yang-Mills model}
\label{sec:noncommutative-yang}
\subsection{Yang-Mills action}
In this section, we will construct an action for an arbitrary noncommutative connection $\hat\nabla$ on a right module $\cM$ over an endomorphism algebra $\aA$ and which is associated to $P$.

We will consider a Riemannian structure on $\aA$, given by a non-degenerate metric $g$ on $\der(\aA)$ and a metric $h$ on $\Oder^{1}(\aA)$ as define in section~\ref{sec:riemannian-structure}.
We have seen that this Riemannian structure naturally define an ordinary connection $\nabla$ on $\End\fE$.
Then, this connection can also be used to decomposed the degrees of freedom of the N.C. connection $\hat\nabla$ in term of a tensorial form $a$ and scalar fields as introduce in the section~\ref{sec:nonc-conn}.
We can associate to this connection a curvature noncommutative $2$-form and consider the minimal action principle based on the functional:
\begin{align*}
  S[\hat\nabla] &=  \Vert\Omega\Vert^{2} =  (\Omega,\Omega) =\fint \langle\Omega,\star\Omega\rangle \ .
\end{align*}
If we use the ordinary connection coming from the metric to decompose the curvature into an horizontal and a vertical part, than this action split into three terms:
  \begin{multline}
    S[\omega] =  \Vert  R(F) + \nabla a  + a^{2} - \varphi(F) \Vert^{2} \\
    + \Vert\nabla_{A}\varphi+[\ta,\varphi] \Vert^{2} 
    + \Vert (\varphi\circ\alpha)^{2} - \varphi\circ\alpha^{2}\Vert^{2} \ .
\label{eq:action-decomp}
  \end{multline}
We can remark that it is essential to use the ordinary connection coming from the metric to decompose the degrees of freedom of the noncommutative connection $\hat\nabla$, otherwise the decomposition of the action would have been a lot more complicated.

This action generalize the action obtain in \cite{dubois-violette:japan99} for the algebra $C^{\infty}(M)\otimes M_{n}(\gC)$, which can be consider as the endomorphism algebra associated to a trivial fiber bundle.
So, we can recover the action found in  \cite{dubois-violette:japan99} by taking a trivial fiber bundle $\fE$ and a gauge potential $A_{\mu}=0$.
The principal difference with the trivial situation correspond to the introduction of a reference connection coming from a Riemannian structure over $\aA$.
This connection is necessary in order to decompose correctly the different local expressions.
However, we must notice that global effects can arise due to the topology of the vector bundle $E$.
So, a careful analysis of the equations of motion and of the vacuums must be perform in this case.

\subsection{Analysis of the vacuums}

For the Euclidian action (with positive signature), the vacuums are given by the following solutions:
\begin{align*}
  & \left\{\begin{aligned}
      &    [\varphi(E_{a}),\varphi(E_{b})] = \varphi([E_{a},E_{b}])\\
      &    \nabla\varphi + [a, \varphi]  =0 \\
            &    R(F) + \nabla a  + a^{2} = \varphi(F) 
          \end{aligned}\right.  &
        & \Rightarrow &
        & 
        \left\{\begin{aligned}
            &\varphi =R_{i} \ \ \text{representation of $\ksu_{n}(\gC)$}\\
            &  d\varphi+ [(R-\varphi)(A)+ a,\varphi]=0\\
            & (R-\varphi)(F)+\nabla a+a^{2}=0
          \end{aligned}\right.
      \end{align*}
      where $A$ is the local gauge potential of $\nabla$.
      
We can try to resolve this system of equations.
We can first remark that there always exists a good global solution which is $\varphi=R$ and $a=0$.
We can remark that for this solution, $\varphi$ is constant.

The existence of other global solutions may depends of the structure of the vector bundle $E$.
To characterize this fact, we can first remark that the potential, the third term in the formula (\ref{eq:action-decomp}), vanish when $\varphi$ is a representation of the Lie algebra $\aA_{0}$. 
In the trivial situation, it was shown in \cite{dubois-violette:japan99} that all this solutions correspond to vacuums and can be map to constant representations of $\ksl_{n}$ by  a gauge transformations. Hence, gauge inequivalent vacuums were classified by inequivalent classes of representations of $\ksl_{n}$.
In the case of a non trivial fiber bundles, we can see that when $\varphi$ is a general representation of $\aA_{0}$, it can no longer be mapped to a constant representation in general (for $\varphi\neq R $ and $\varphi\neq 0$).
This simply mean that a gauge transformation on a noncommutative connection can not compensate the variations of fields due to changing of charts.
This phenomenon also mean that general configurations which have a zero potential energy will be necessary non constant and so, most of the time,  they will not describe vacuums.
We have yet to discuss the case of the trivial representation $\varphi=0$, $a=0$.
We can see that this configuration correspond to a vacuum if and only if the reference curvature $2$-form $F$ is zero.
So, if the fiber bundle $E$ does not admit connections with vanishing curvature, this configuration will not correspond to a vacuum.

We can conclude from this analysis that in the case of a non trivial fiber bundle, it can happen that some vacuums of the trivial situation disappear.
The only vacuum of the trivial situation which remain to be a vacuum is the configuration $\varphi=R, a=0$ which correspond to the massive sector in the Higgs mechanism picture.
We can so naively say that the topology of the space-time manifold $M$ seems disturb the Higgs mechanism in such a model of symmetry breaking and that the non trivial structure of the vector bundle furnish a kind of selection rule in the vacuums.


\subsection*{Acknowledgments}
I would like to thank M. Dubois-Violette and T. Masson for interesting discussions about various parts of this work.

\newpage
\appendix

\section*{Appendix}
\section{Levi-Cività connection}

\subsection{Definitions}
It is possible to introduce the concept of linear connection~\cite{dubo-mich:95} on $\der(\aA)$ and 
to associate to any Riemannian structure a unique linear connection without torsion which left the metric invariant.
We call such a connection a Levi-Cività connection.

We precise this notions in the following definitions:

\begin{defn}[Linear connection]
  A linear connection is a connection on the $Z(A)$-module  $\der(\aA)$, \textit{i.e.} an application:
  \begin{align*}
\der(\aA) &\to \End(\der(\aA))\\
X  &\mapsto    D_{\cX}
  \end{align*}
which satisfy
\begin{align*}
  D_{\cX}(f \cY) &= \cX(f) \cY + f D_{\cX}(\cY) &
  D_{f\cX}(\cY) &= f D_{\cX}(\cY) &
  \forall \cX,\cY \in \der(\aA),  \forall f\in Z(\aA) \ .
\end{align*}

The application $D$ can also be view as an application
\begin{align*}
 \der(\aA)&\to \Oder(\aA,\der(\aA))
\end{align*}
\end{defn}

\begin{defn}[Torsion]
  The torsion $T^{D}$ associate to a linear connection $D$ is the  noncommutative tensorial $2$-form  $T^{D} \in  \Oder^{2}(\aA,\der(\aA))$ defined by the formula:
  \begin{align*}
    T^{D}(\cX,\cY) &= D_{\cX}(\cY) - D_{\cY}(\cX) - [\cX,\cY] \ .
  \end{align*}
\end{defn}

\begin{defn}[Levi-Cività connection]
  The Levi-Cività connection associated to a metric $g$ on $\der(\aA)$ is the unique linear connection without trosion ($T^{D}=0$) which left the metric $g$ invariant:
  \begin{align*}
    \cX(g(\cY,\cZ)) &=  g(D_{\cX}\cY, \cZ) + g(\cY,D_{\cX}\cY)  \ ,
  \end{align*}  
  Then, the Levi-Cività connection is defined by the following relations:
  \begin{multline}
    2 g(D_{\cX}\cY, \cZ) = \cX (g(\cY,\cZ)) + \cY(g(\cX,\cZ)) -\cZ(g(\cX,\cY)) \\
    + g([\cX,\cY],\cZ) + g([\cZ,\cY],\cX) + g([\cZ,\cX],\cY) \ .
  \end{multline}
\end{defn}

\subsubsection{Local expressions and Christoffel symbols}

We consider a metric $g$ on $\der(\aA)$ and a metric $h$ on  $\Oder(\aA)$ which define a Riemannian structure on $\aA$.
Let $D$ be the Levi-Cività connection canonicaly associated to it.
The expressions of the Levi-Cività connection in the local basis of derivations $(\nabla_{ \mu}, \ad_{E_{a}})$ will define the Christoffel symbols.
With the notations take in section~\ref{sec:local-expr-metr}, we find the following expressions for the covariant derivatives:
\begin{align*}
  D_{\nabla_{\mu}}\nabla_{\nu} &= \frac{1}{2} \ad_{F_{\mu \nu}} + \Gamma_{\mu \nu}^{\sigma} \nabla_{\sigma} + \Gamma_{\mu \nu}^{d} \ad_{E_{d}} \\
  D_{\nabla_{\mu}}\ad_{E_{b}} &= \ad_{\nabla_{\mu}E_{b}} + \Gamma_{\mu b}^{\sigma} \nabla_{\sigma} + \Gamma_{\mu b}^{d} \ad_{E_{d}} \\
  D_{\ad_{E_{a}}} \nabla_{\nu} &= \Gamma_{a \nu}^{\sigma} \nabla_{\sigma} + \Gamma_{a \nu}^{d} \ad_{E_{d}} \\
  D_{\ad_{E_{a}}} \ad_{E_{b}} &= \frac{1}{2} \ad_{[E_{a},E_{b}]}  + \Gamma_{a b}^{\sigma} \nabla_{\sigma} + \Gamma_{a b}^{d} \ad_{E_{d}} \ .
\end{align*}
where the coefficients $\Gamma_{AB}^{C}$ are the Christoffel symbols.
They are defined by the following formulas:
\begin{align*}
  \Gamma_{\mu \nu}^{\sigma} &= \frac{1}{2}h^{\sigma \rho}(\partial_{\mu}g^{M}_{\nu \rho} + \partial_{\nu}g^{M}_{\mu \rho} - \partial_{\rho}g^{M}_{\mu \nu})&
  \Gamma_{\mu \nu}^{d}&=  0 \\
  \Gamma_{\mu b}^{\sigma} &=\frac{1}{2}h^{\sigma \rho} g_{eb} F_{\rho \mu}^{e} &
  \Gamma_{\mu b}^{d}&= \frac{1}{2}h_{\Int}^{dc} \nabla_{\mu} g_{b c}    \\
  \Gamma_{a \nu}^{\sigma} &= \frac{1}{2}h^{\sigma \rho} g_{ea} F_{\rho \nu}^{e} &
  \Gamma_{a  \nu}^{d}&= \frac{1}{2}h_{\Int}^{dc} \nabla_{\nu} g_{a c}   \\
  \Gamma_{a b}^{\sigma} &= -\frac{1}{2} h^{\sigma \rho} \nabla_{\rho} g_{ab}&
  \Gamma_{a b}^{d}&= \frac{1}{2} h_{\Int}^{dc} L_{\ad_{E_{c}}} g_{ab}
\end{align*}
where we have used the notations:
\begin{align*}
  &  L_{\ad_{E_{c}}} g_{ab} =  (L_{\ad_{E_{c}}} g_{\Int})(E_{a},E_{b}) 
  = -C_{ca}^{e}g_{eb} -   C_{cb}^{e}g_{ae} \\
  &  \nabla_{\mu}E_{a} = [A_{\mu}, E_{a}]  
  = A_{\mu}^{b}C_{ba}^{c} E_{c} \\
  & \nabla_{\mu}g_{ab}=  (L_{\nabla_{\mu}}g_{\Int})(E_{a},E_{b}) 
  = \partial_{\mu}g_{ab} - A_{\mu}^{e}C_{ea}^{f}g_{fb} - A_{\mu}^{e}C_{eb}^{f}g_{af}\\
  &F_{\mu \nu} = [\nabla_{\mu},\nabla_{\nu}] = \partial_{\mu} A_{\nu} - \partial_{\nu} A_{\mu} + [A_{\mu}, A_{\nu}] \ .
\end{align*}

This linear connection has the same decomposition than the Levi-Cività connection defined in a Kaluza-Klein theory on a principal fiber bundle with structure group $SU(n)$ (see \cite{kerner:81,kerner:88,Gior-kern:88}) and the calculations of quantities like curvature are identical than in the present context. 
The scalar curvature can be used to define an Einstein-Hilbert action for the algebra $\aA$. 
This action can be decomposed into three parts, corresponding to vertical and horizontal degrees of freedom, as we have done for the Yang-Mills type action for noncommutative connections.
One part will be the ordianry Einstein-Hilbert action associated to the metric $g^{M}$ on the base manifold.
An other part will be a Yang-Mills action for the ordinary connection $\nabla$ associated to the metric (see section \ref{sec:decomp-metr-refer-1}).
The last part is an action with a quartic potential for scalar fields. This scalar fields correspond to the vertical degrees of freedom of the metric.
We don't reproduce here the calculations because they are identical to the one made in~\cite{kerner:81,kerner:88,Gior-kern:88} in the context of a Kaluza-Klein theory.



\newpage

\bibliography{biblio_articles,biblio_livres,gnc,K-theory}
\bibliographystyle{utphys}

\end{document}